\documentclass[aps,pre,twocolumn,groupedaddress,superscriptaddress,showpacs]{revtex4-1}

\usepackage{graphicx}
\usepackage{epstopdf}
\usepackage{amsmath}
\usepackage{amsthm}
\usepackage{amsfonts}
\usepackage{subfigure}
\usepackage{hhline}
\usepackage[miktex]{gnuplottex}
\usepackage{xcolor}
\usepackage{color} 
\usepackage[normalem]{ulem}

\newcommand{\avrg}[1]{\left\langle #1 \right\rangle}
\newcommand{\nn}{\nonumber \\}
\newcommand{\nnn}{\nonumber}

\begin{document}

\title{
Towards a generalization of information theory for hierarchical partitions
}

\author{Juan Ignacio Perotti}
\affiliation{Facultad de Matem\'atica, Astronom\'ia, F\'isica y Computaci\'on, Universidad Nacional de C\'ordoba, Ciudad Universitaria, C\'ordoba, Argentina.}
\affiliation{Instituto de F\'isica Enrique Gaviola (IFEG-CONICET), Ciudad Universitaria, C\'ordoba, Argentina.}
\email[E-mail: ]{juan.perotti@unc.edu.ar}

\author{Nahuel Almeira}
\affiliation{Facultad de Matem\'atica, Astronom\'ia, F\'isica y Computaci\'on, Universidad Nacional de C\'ordoba, Ciudad Universitaria, C\'ordoba, Argentina.}
\affiliation{Instituto de F\'isica Enrique Gaviola (IFEG-CONICET), Ciudad Universitaria, C\'ordoba, Argentina.}

\author{Fabio Saracco}
\affiliation{IMT School for Advanced Studies Lucca, Piazza San Francesco 19, I-55100, Lucca, Italy}

\date{\today}

\begin{abstract}
Complex systems often exhibit multiple levels of organization covering a wide range of physical scales, so the study of the hierarchical decomposition of their structure and function is frequently convenient.
To better understand this phenomenon, we introduce a generalization of information theory that works with hierarchical partitions.
We begin revisiting the recently introduced Hierarchical Mutual Information (HMI), and show that it can be written as a level by level summation of classical conditional mutual information terms.
Then, we prove that the HMI is bounded from above by the corresponding hierarchical joint entropy.
In this way, in analogy to the classical case, we derive hierarchical generalizations of many other classical information-theoretic quantities.
In particular, we prove that, as opposed to its classical counterpart, the hierarchical generalization of the Variation of Information is not a metric distance, but it admits a transformation into one.
Moreover, focusing on potential applications of the existing developments of the theory, we show how to adjust by chance the HMI.
We also corroborate and analyze all the presented theoretical results with exhaustive numerical computations, and include an illustrative application example of the introduced formalism.
Finally, we mention some open problems that should be eventually addressed for the proposed generalization of information theory to reach maturity. 
\end{abstract}

\maketitle

\section{INTRODUCTION}

Information theory plays an important role in physics at the fundamental, theoretical, and application levels~\cite{hawking1975particle,
mezard2009information,
witten2020mini,
newman2020improved}.
In particular, Jaynes~\cite{Jaynes1957} showed how to derive the ensembles of statistical physics from information theory, simply considering the energy of the system as the available information. The approach of Jaynes found many applications.
For example, Park and Newman~\cite{Park2004} extended it to complex networks, providing an unbiased framework for their analysis, which was later refined to study online social networks, the international trade network, and financial networks~\cite{Cimini2018a}.

The Renormalization Group theory of statistical mechanics reveals how information aggregates through a wide range of physical scales giving rise to emergent phenomena~\cite{zinnjustin2007phase}. 
Analogously, in the context of complex systems, multiple levels of organization often emerge, and their study through the hierarchical decomposition of their structure and function is generally convenient.
The study of complex phenomena through hierarchical representations has found several applications~\cite{ravasz2003hierarchical,
guimera2003self,
song2005self,
zhou2006hierarchical,
muchnik2007self,
helbing2015saving,
jalili2017information,
shekhtman2018percolation,
garcia2018multiscale,
lee2018hierarchical,
gates2019element,
salichos2014novel,
bassolas2019hierarchical}.
Certainly, the generalization of information theory to hierarchical representations is an inquiring research topic and our paper contributes to its development.

Most results in classical information theory could be summarized in a few definitions~\cite{cover2006elements}.
For instance, many information-theoretic quantities can be derived from the definition of mutual information.
This is useful for the generalization of classical information theory.
A paradigmatic case is found in quantum mechanics~\cite{witten2020mini}, where entropies can be redefined as operators over a Hilbert space instead of functionals over probability distributions.
The quantum mechanical generalization of information theory has influential consequences.
For example, despite the fact that conditional probabilities operate differently in quantum mechanics and classical physics, many results in classical information-theory remain true in the quantum case.
In a sense, probabilities only provide a particular form of encoding information about partitions, and information theory goes beyond probability theory.
Since hierarchical partitions constitute a generalization of partitions, the recently introduced Hierarchical Mutual Information (HMI)~\cite{perotti2015hierarchical} conveys a natural starting point for a corresponding generalization of information theory.
This is the approach we  decided to follow.

Finding appropriate hierarchical decomposition of the structure and function of a system is a challenging issue~\cite{pardo2007extracting,crutchfield1994calculi,rosvall2011multilevel,queyroi2013assesing,peixoto2014hierarchical,rosvall2014memory,tibely2016comparing,grauwin2017identifying}.
Here, to detect statistically significant hierarchical decomposition, the adequate comparison of hierarchical structures is of crucial importance.
Several comparison methods already exist, including tree-edit distance methods~\cite{bille2005survey,zhang2009split,queyroi2015suppression}, ad-hoc methods~\cite{sokal1962comparison,fowlkes1983method,gates2019element}, and information-theoretic methods~\cite{tibely2014extracting,perotti2015hierarchical}.
In this regard, the HMI is a generalization of the traditional Mutual Information (MI)~\cite{danon2005comparing} to the hierarchical case, and it has already found successful applications in the comparison of hierarchical community structures~\cite{kheirkhahzadeh2016efficient,yang2017hierarchical}.
Notice, however, that without an appropriate theoretical background, the HMI can be easily criticized as a similarity measure~\cite{gates2019element}.
For example, a well-known problem of the non-hierarchical mutual information is the necessity of a null-model adjustment~\cite{meila2007comparing,vinh2009information,zhang2015evaluating,newman2020improved}.
As we show in this work, the problem persists in the hierarchical case, but, thanks to the theoretical development we provide, we fix this inconvenience by rendering an adjusted version of the HMI.
Moreover, we also derive a hierarchical information-theoretic metric distance~\cite{meila2007comparing}, enabling a potential geometrization of the space of hierarchical partitions.
We also study the numerical properties of the introduced similarity and distance quantities, including a simple example application in hierarchical clustering.

Let us summarize the content of the forthcoming sections.
In Sec.~\ref{sec:theory}, we introduce some preliminary definitions and revisit the HMI.
In Sec.~\ref{sec:results}, we present the main results.
In Sec.~\ref{sec:results}A, we prove some fundamental properties of the HMI.
In Sec.~\ref{sec:results}B, we use the HMI to introduce other information-theoretic quantities for hierarchical-partitions.
In particular, we study the metric properties of the Hierarchical Variation of Information (HVI) and introduce a metric distance.
We also define and study the statistical properties of an Adjusted HMI (AHMI).
In Sec.~\ref{sec:results}C, we show a simple application of the introduced framework.
In Sec.~\ref{sec:discussion} we discuss some important consequences deriving from the presented results and discuss corresponding opportunities for future works.
Finally, in Sec.~\ref{sec:conclusions} we provide a summary of the contributions.

\section{\label{sec:theory} THEORY}

\subsection{Preliminary definitions}

Let $T$ denote a directed rooted tree.
We say that $t\in T$ when $t$ is a node of $T$.
Let $T_t$ be the set of children of node $t\in T$.
If $T_t=\emptyset$ then $t$ is a leaf of $T$. 
Otherwise, it is an internal node of $T$.
Let $\ell_t$ denote the depth or topological distance between $t$ and the root of $T$.
In particular, $\ell_t=0$ if $t$ is the root.
Let $T_{\ell}$ be the set of all nodes of $T$ at depth $\ell$.
Clearly $T_{\ell+1}=\cup_{t\in T_{\ell}} T_t$.
Let $T^t$ be the sub-tree obtained from $t$ and its descendants in $T$.

\begin{figure}
	\includegraphics[angle=-90,scale=.9]{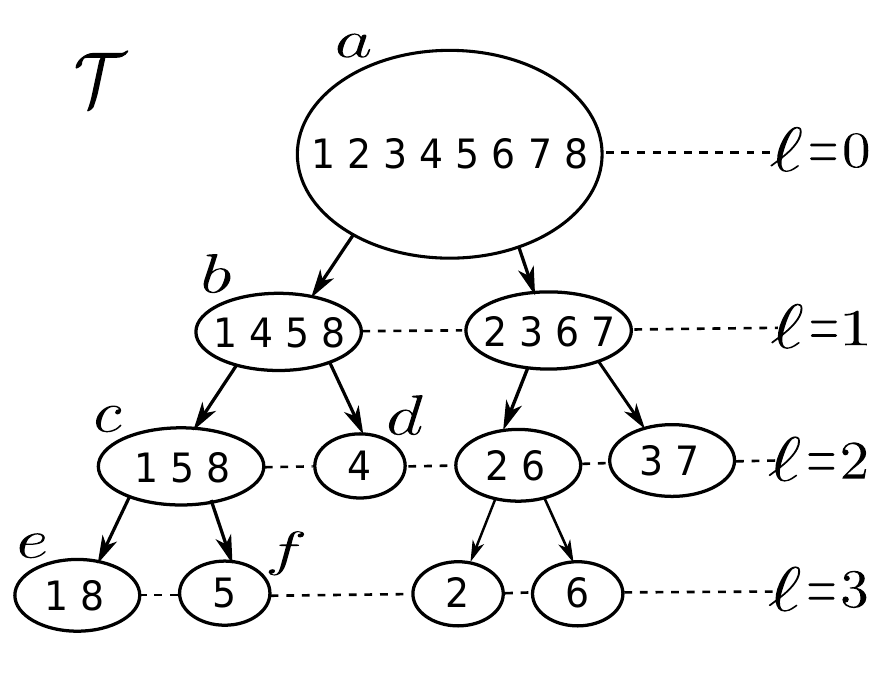}
	\caption{
		\label{fig:1}
		Schematic representation of a hierarchical-partition $\mathcal{T}$ of the universe $U=\{1,2,...,8\}$ with root $a$, 5 internal nodes including $a,b$ and $c$, and 6 leaves including $d,e$, and $f$.
		Some leaves may contain more than one element, e.g. $U_e=\{1,8\}$.
		Different leaves may exist at different depths $\ell$. 
		For instance, leaf $d$ is at depth $\ell=2$ while leaf $f$ is at depth $\ell=3$.
		The sub-tree $T^b$ contains the nodes $b,c,d,e$, and $f$.
		The set $T_b$ contains the children $c$ and $d$ of $b$.
	}
\end{figure}

A {\em hierarchical-partition} $\mathcal{T}:=\{U_t:t\in T\}$ of the {\em universe} $U:=\{1,...,n\}$, the set of the first $n$ natural numbers, is defined in terms of a rooted tree $T$ and corresponding subsets $U_t\subset U$ satisfying
\begin{itemize}
	\item[{\em i)}] $\cup_{t'\in T_t} U_{t'} = U_t$ for all non-leaf $t$, and
	\item[{\em ii)}] $U_{t'}\cap U_{t''} = \emptyset$ for every pair of different $t',t'' \in T_t$.
\end{itemize}
For every non-leaf $t$, the set $\mathcal{T}_t:= \{U_{t'}:t'\in T_t \}$ represents a partition of $U_t$, and $\mathcal{T}_{\ell}:=\{U_t:t\in T_{\ell}\}$ 
is the ordinary partition of $U$ determined by $\mathcal{T}$ at depth $\ell$.
Furthermore, $\mathcal{T}^t:=\{U_{t'}:t'\in T^t\}$ is the hierarchical-partition of the universe $U_t$ determined by the tree $T^t$ of root $t$.
See Fig.~\ref{fig:1} for a schematic representation of a hierarchical-partition of the universe $U=\{1,2,...,8\}$.

\subsection{The Hierarchical Mutual Information}

The HMI~\cite{perotti2015hierarchical} between two hierarchical-partitions $\mathcal{T}$ and $\mathcal{S}$ of the same universe $U$ reads
\begin{eqnarray}
\label{eq:1}
I(\mathcal{T};\mathcal{S})
&:=&
I(\mathcal{T}^{t_0};\mathcal{S}^{s_0})
\end{eqnarray}
where $t_0$ and $s_0$ are the roots of trees $T$ and $S$, respectively.
Here,
\begin{eqnarray}
\label{eq:2}
I(\mathcal{T}^t;\mathcal{S}^s)
&:=&
I(T_t;S_s|ts)
+
\sum_{\substack{t'\in T_t\\s'\in S_s}}
P(t's'|ts)
I(\mathcal{T}^{t'};\mathcal{S}^{s'})
\end{eqnarray}
is a recursively defined expression for every pair of nodes $t\in T$ and $s\in S$ with the same depth $\ell_t = \ell_s$.
The probabilities in $P(t's'|ts)=P(t's'ts)/P(ts)$ are ultimately defined from
$P(t's'ts) := |U_{t'} \cap U_{s'} \cap U_{t} \cap U_{s}|/|U|$ and the convention $0/0 = 0$.
The quantity
\begin{eqnarray}
\label{eq:3}
I(T_t;S_s|ts)
&:=&
H(T_t|ts)
+
H(S_s|ts)
-
H(T_t,S_s|ts)
,
\end{eqnarray}
represents a mutual information between the standard partitions $\mathcal{U}_t$ and $\mathcal{U}_s$ restricted to the subset $U_t\cap U_s$ of the universe $U$, and is defined in terms of the three entropies
\begin{eqnarray}
\label{eq:4}
H(T_t|ts)
:=
\sum_{t'\in T_t}
-
P(t'|ts)
\ln
P(t'|ts),
\end{eqnarray}
\begin{eqnarray}
\label{eq:5}
H(T_s|ts)
:=
\sum_{s'\in S_s}
-
P(s'|ts)
\ln
P(s'|ts)
\end{eqnarray}
and
\begin{eqnarray}
\label{eq:6}
H(T_t,S_s|ts)
&:=&
\sum_{\substack{t'\in T_t\\s'\in S_s}}
-
P(t's'|ts)
\ln
P(t's'|ts)
\end{eqnarray}
where the convention $0\ln 0=0$ is adopted.
For details on how to compute these quantities, please check our code~\cite{perotti2019hit_gitcode}.

\section{\label{sec:results} RESULTS}

For simplicity, we consider hierarchical-partitions $\mathcal{T}$ and $\mathcal{S}$ with all leaves at depths $\ell=L>0$.
The results can be easily generalized to trees with leaves at different depths at the expense of using more complicated notation.

\subsection{Properties of the HMI}

It is convenient to begin rewriting the hierarchical mutual information in the following alternative form, which is more convenient for our purposes (see Appendix~\ref{app:A} for a detailed derivation)
\begin{eqnarray}
\label{eq:7}
I(\mathcal{T};\mathcal{S})
&=&
I(\mathcal{T}^{t_0};\mathcal{S}^{s_0})
\\
&=&
\sum_{\ell=0}^{L-1}
\sum_{t\in T_{\ell},s\in S_{\ell}}
P(t s)
I(T_t;S_s|ts)
\nn
&=&
\sum_{\ell=0}^{L-1}
I(T_{{\ell}+1};S_{{\ell}+1}|T_{\ell},S_{\ell})
,
\nnn
\end{eqnarray}
where $P(t s):=P(t s|t_0 s_0)$ and, 
as the reader can see, we rewrote the HMI as a level by level summatory of classical (i.e. non-hierarchical) conditional MIs.
This is useful because it allows us to study the difference between two hierarchical partitions under a level by level basis~\cite{perotti2015hierarchical}.
Other methods such as edit distances or ad-hoc methods~\cite{gates2019element} do not offer the possibility of studying the contribution of each vertex or level of the hierarchy in an independent way.
In particular, non-informative vertices or levels composed of trivial partitions produce null contributions within the HMI, which is convenient for the comparison of hierarchical partitions.
Later, in Sec.~\ref{sec:III:C}, we show the advantages of the properties of the HMI in the analysis of an illustrative application.

Starting from Eq.~\ref{eq:7}, we prove the following property of the HMI (see Appendix~\ref{app:B} for a detailed derivation)
\begin{equation}
\label{eq:8}
0
\leq
I(\mathcal{T};\mathcal{S})
\leq 
I(\mathcal{T};\mathcal{T})
.
\end{equation}
In other words, this result states that the HMI between two arbitrary hierarchical-partitions $\mathcal{T}$ and $\mathcal{S}$ of the same universe $U$ is smaller or equal to the mutual information between $\mathcal{T}$ and itself (or analogously between $\mathcal{S}$ and itself) mimicking in this way an analogous property that holds for the classical mutual information~\cite{cover2006elements}.

Now we exploit the result of Eq.~\ref{eq:8} to show that the HMI can be properly normalized.
Namely, if $M(x,y)$ is any generalized mean~\cite{bullen2003handbook} (like 
the arithmetic-mean $M(x,y)=(x+y)/2$, 
the geometric-mean $M(x,y)=\sqrt{xy}$, 
the max-mean $M(x,y)=\max(x,y)$
or 
the min-mean $M(x,y)=\min(x,y)$) then the Normalized HMI (NHMI) 
\begin{eqnarray}
\label{eq:9}
i(\mathcal{T};\mathcal{S})
:=
\frac{
I(\mathcal{T};\mathcal{S})
}{
M(
H(\mathcal{T}),
H(\mathcal{S})
)
}
\end{eqnarray}
satisfies $0\leq i(\mathcal{T};\mathcal{S}) \leq 1$.
Both inequalities follow from Eq.~\ref{eq:8}.

\subsection{Deriving other information-theoretic quantities}

\begin{figure}
	\includegraphics{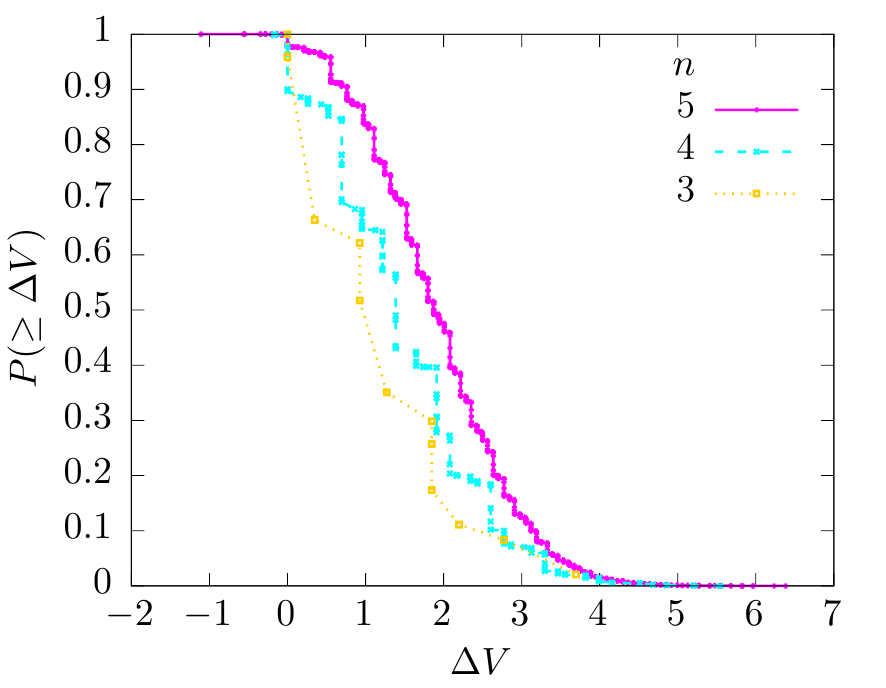}
	\caption{
		\label{fig:2}
		(Color online)
		Complementary cumulative distribution of inequalities $\Delta V$ for the Hierarchical Variation of Information $V$ for different hierarchy sizes $n$.
		Negative values exist, breaking triangular inequality, although most of them are positive and over a wider range.
	}
\end{figure}

Given the HMI, hierarchical versions of other information-theoretic quantities can be obtained by following the rules of the standard classical case.
For example, the Hierarchical Entropy (HE) of a hierarchical-partition $\mathcal{T}$ can be defined as
\begin{eqnarray}
\label{eq:10}
H(\mathcal{T})
&:= &
I(\mathcal{T};\mathcal{T})
\\
&=&
\sum_{\ell=0}^{L-1}
H(T_{\ell+1}|T_{\ell})
\nn
&= &
\sum_{\ell=0}^{L-1}
\bigg(
H(T_{\ell+1},T_{\ell})
-
H(T_{\ell})
\bigg)
\nn
&= &
\sum_{\ell=0}^{L-1}
\bigg(
H(T_{\ell+1})
-
H(T_{\ell})
\bigg)
\nn
&= &
H(T_L)
\nnn
\end{eqnarray}
where we used that $H(T_{\ell+1},T_{\ell}) = H(T_{\ell+1})$ (see Eq.~\ref{eq:D6}).
Similarly, we can write down the Hierarchical Joint Entropy (HJE) as
\begin{eqnarray}
\label{eq:11}
H(\mathcal{T},\mathcal{S})
:=
H(\mathcal{T})
+
H(\mathcal{S})
-
I(\mathcal{T};\mathcal{S})
\end{eqnarray}
and the Hierarchical Conditional Entropy (HCE) as
\begin{eqnarray}
\label{eq:12}
H(\mathcal{T}|\mathcal{S})
&:=&
H(\mathcal{T},\mathcal{S})
-
H(\mathcal{S})
\\
&=&
H(\mathcal{T})
-
I(\mathcal{T};\mathcal{S})
.
\nnn
\end{eqnarray}
Furthermore, we can define the Hierarchical Variation of Information (HVI) as
\begin{eqnarray}
\label{eq:13}
V(\mathcal{T};\mathcal{S})
&:=&
H(\mathcal{T}|\mathcal{S})
+
H(\mathcal{S}|\mathcal{T})
\\
&=&
H(\mathcal{T})
+
H(\mathcal{S})
-
2I(\mathcal{T};\mathcal{S})
\nn
&=&
H(\mathcal{T},\mathcal{S})
-
I(\mathcal{T};\mathcal{S})
.
\nnn
\end{eqnarray}
Because of Eq.~\ref{eq:8}, the properties
$H(\mathcal{T},\mathcal{S})\geq H(\mathcal{T}) \geq H(\mathcal{T}|\mathcal{S})\geq 0$
and
$V(\mathcal{T};\mathcal{S})\geq 0$ follow,
generalizing corresponding properties of the classical case.
Unfortunately, we found counter-examples violating the triangle inequality for the HVI, failing to generalize its classical counterpart in this particular sense~\cite{meila2007comparing}.
For instance, for the hierarchical-partitions
$\mathcal{T}=[[[1,2],[3]],[4]]$,
$\mathcal{S}=[[2],[[3],[1,4]]]$
and
$\mathcal{R}=[[1],[2],[[3],[4]]]$,
we find
$V(\mathcal{T};\mathcal{S})
+
V(\mathcal{S};\mathcal{R})
- 
V(\mathcal{T};\mathcal{R})
\approx 
-0.17$, which is a negative quantity.
It is important to remark, however,
that the violation of the triangular inequality is relatively weak.
For instance, for $n=4$ the maximum difference is found to be
$\approx 
5.55$ for $\mathcal{T}=[[[1],[2]],[[3],[4]]]$,
$\mathcal{S}=[[[1],[3]],[[2],[4]]]$
and
$\mathcal{R}=[[[1],[2]],[[3],[4]]]$, which is significantly larger than $0.17$.
In fact, as shown in 
Fig.~\ref{fig:2} where the complementary cumulative distribution of differences
\begin{eqnarray}
\label{eq:14}
\Delta V(\mathcal{T},\mathcal{S},\mathcal{R})
&:=&
V(\mathcal{T};\mathcal{S})
+
V(\mathcal{S};\mathcal{R})
- 
V(\mathcal{T};\mathcal{R})
\end{eqnarray}
is plotted for all 
$\mathcal{T}$, $\mathcal{S}$ and $\mathcal{R}$ without repeating the symmetric cases 
$\Delta V(\mathcal{T},\mathcal{S},\mathcal{R})$
and
$\Delta V(\mathcal{R},\mathcal{S},\mathcal{T})$,
and for different sizes $n$, the overall contribution of the negative values is small, not only in magnitude but also in probability.
Results for larger values of $n$ are not included since the number of triples $(\mathcal{T},\mathcal{S},\mathcal{R})$
grows quickly with $n$, turning impractical their exhaustive computation.
See Appendix~\ref{app:C} for how to generate all possible hierarchical-partitions for a given $n$.

\begin{figure}
	\includegraphics{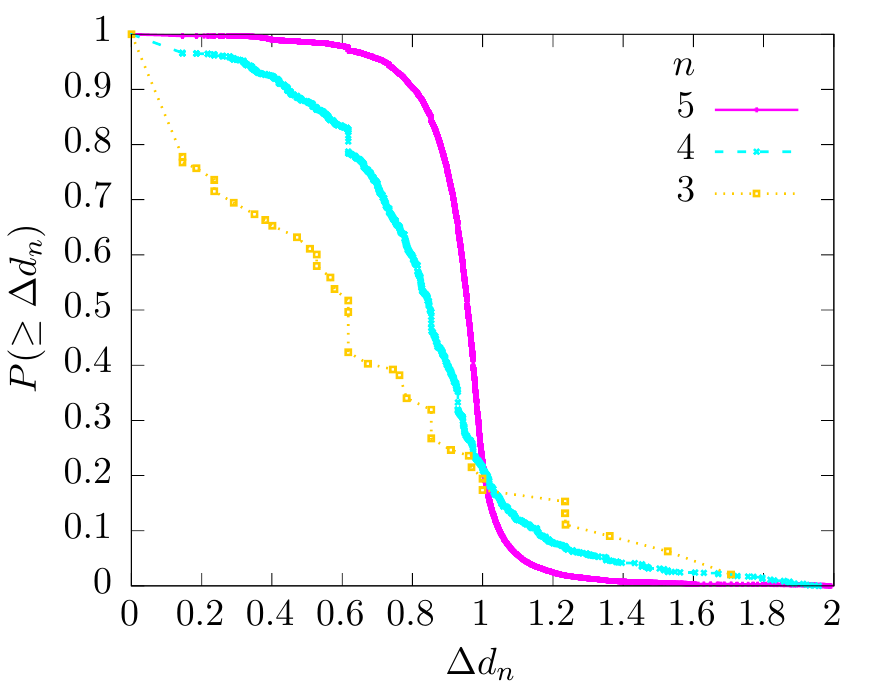}
	\caption{
		\label{fig:3}
		(Color online)
		Complementary cumulative distribution of inequalities $\Delta d_n$ for the distance metric $d_n$ derived from the Hierarchical Variation of Information $V$ for different hierarchy sizes $n$.
		All values are non-negative in agreement with the theory.
	}
\end{figure}

Although the HVI fails to satisfy the triangular inequality, the transformation
\begin{eqnarray}
\label{eq:15}
d_n(\mathcal{T},\mathcal{R})
&=&
1-e^{-n\frac{\ln 2}{2} V(\mathcal{T},\mathcal{R})}
\end{eqnarray} 
of $V$ does it (see Appendix~\ref{app:D} for a detailed proof).
In other words, $d_n$ is a distance metric, so the geometrization of the set of hierarchical-partitions is possible.
We confirm this in Fig.~\ref{fig:3} by running computations analogous to those of Fig.~\ref{fig:2} but for $\Delta d_n$ instead of $\Delta V$.
Notice however that the distance metric $d_n$ is non-universal, because it depends on $n$.
In fact, for $n\to \infty$ it holds $d_{n}(\mathcal{T};\mathcal{S}) \to 1-\delta_{\mathcal{T},\mathcal{S}}$ 
which is a trivial distance metric (known as the discrete metric) that can only distinguish between equality and non-equality.
These properties follow because, for fixed-size $n$, the non-zero $V$'s are bounded from below by a finite positive quantity that tends to zero when $n\to \infty$.
We also remark that other concave growing functions besides that of Eq.~\ref{eq:15} (or more specifically Eq.~\ref{eq:D1}) can be used to obtain essentially the same result\textcolor{magenta}{, i.e. }
a distance metric.

Although the classical VI is a distance metric---which is a desirable property for the quantitative comparison of entities---it also presents some limitations~\cite{fortunato2016community}.
Hence, besides the HVI, the HMI, and the NHMI, it is convenient to consider other information-theoretic alternatives for the comparison of hierarchies.
This is the case of the Adjusted Mutual Information (AMI)~\cite{vinh2009information}, which is devised to compensate for the biases that random coincidences produce on the NMI, and which we generalize into the hierarchical case by following the original definition recipe
\begin{eqnarray}
\label{eq:16}
A(\mathcal{T};\mathcal{S})
&:=&
\frac{
I(\mathcal{T};\mathcal{S})
-
\avrg{I(\mathcal{T};\mathcal{S})}
}{
M(H(\mathcal{T}),H(\mathcal{S}))
-
\avrg{I(\mathcal{T};\mathcal{S})}
}
.
\end{eqnarray}
We called the generalization, the Adjusted HMI (AHMI).
The definition of the AHMI requires the definition of a hierarchical version (EHMI)
\begin{eqnarray}
\label{eq:17}
\avrg{I(\mathcal{T};\mathcal{S})}
&:=&
\sum_{\mathcal{R},\mathcal{Q}}
P(\mathcal{R},\mathcal{Q}|\mathcal{T},\mathcal{S})
I(\mathcal{R};\mathcal{Q})
\end{eqnarray}
of the Expected Mutual Information (EMI)~\cite{vinh2009information}.
Here, the distribution
$P(\mathcal{R},\mathcal{Q}|\mathcal{T},\mathcal{S})$
represents a reference null model for the randomization of
a pair of hierarchical-partitions.
Like in the original classical case~\cite{vinh2009information},
we define the distribution in terms of the well-known permutation model.
It is important to remark, however, that other alternatives for the classical case have been recently proposed~\cite{newman2020improved}.

To describe the permutation model, let us first introduce some definitions.
A permutation $\tau$ is a bijection $e\leftrightarrow \tau(e)$ over $U$.
We can define
$\tau \mathcal{T}:=\{\tau U_t:t\in T\}$ 
as the hierarchical-partition of the permuted elements where
$\tau U_t:=\{\tau(e):e\in U_t\}$ for all $t\in T$.
In this way, $\tau \mathcal{T}_{\ell}:=\{\tau U_r:r\in T_{\ell}\}$ becomes the partition emerging at depth $\ell$ obtained from the permuted elements.

Now we are ready to define the permutation model for hierarchical-partitions.
Consider a pair of permutations $\tau$ and $\sigma$ over $U$ acting on corresponding hierarchical-partitions $\mathcal{T}$ and $\mathcal{S}$.
The permutation model is defined as
\begin{eqnarray}
\label{eq:18}
P(\mathcal{R},\mathcal{Q}|\mathcal{T},\mathcal{S})
&:=&
\frac{1}{(n!)^2}
\sum_{\tau,\sigma}
\delta_{\mathcal{R},\tau \mathcal{T}}
\delta_{\mathcal{Q},\sigma \mathcal{S}}
\end{eqnarray}
In this way, Eq.~\ref{eq:17} can be written as
\begin{eqnarray}
\label{eq:19}
\avrg{I(\mathcal{T};\mathcal{S})}
&=&
\frac{1}{(n!)^2}
\sum_{\tau,\sigma}
I(\tau \mathcal{T};\sigma \mathcal{S})
\\
&=&
\frac{1}{n!}
\sum_{\rho}
I(\rho \mathcal{T};\mathcal{S})
\nnn
\end{eqnarray}
where the simplification $\rho=\tau \sigma^{-1}$ can be used because the labeling of the elements in $U$ is arbitrary.

The exact computation of Eq.~\ref{eq:19} is expensive, even if the expressions are written in terms of contingency tables and corresponding generalized multivariate two-way hypergeometric distributions.
This is because, at variance with the classical case, independence among random variables is compromised.
Hence, we approximate the EHMI by sampling permutations $\rho$ until the relative error of the mean falls below $0.01$.

\begin{figure}
	\includegraphics{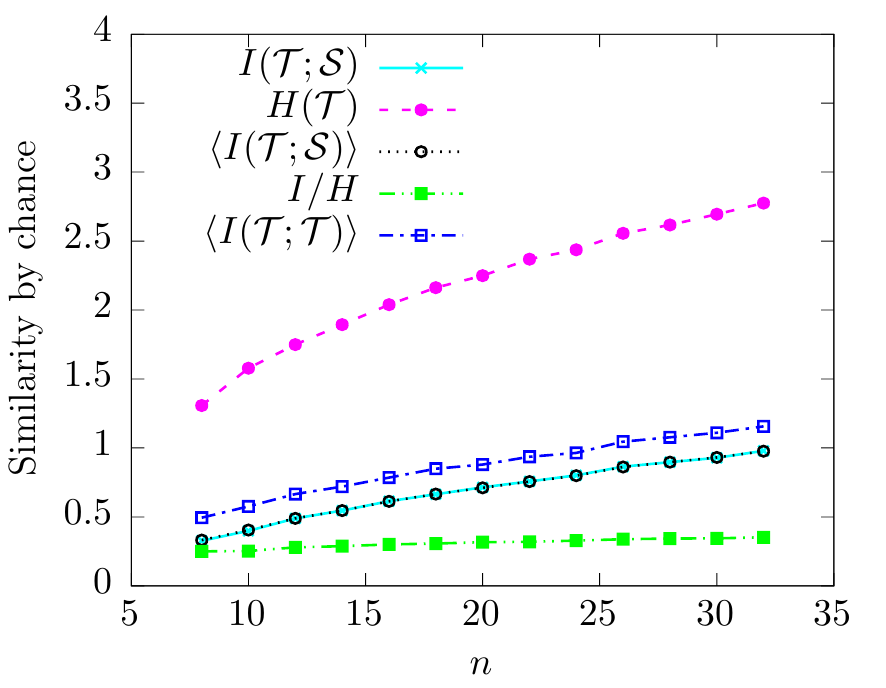}
	\caption{
		\label{fig:4}
		(Color online)
		How similarity by chance affects the Hierarchical Mutual Information $I$.
		In cyan crosses, values of $I$ averaged by sampling pairs of randomly generated hierarchical-partitions $\mathcal{T}$ and $\mathcal{S}$ of the universe with $n$ elements.
		In solid magenta circles, the average hierarchical entropy over the sampled $\mathcal{T}$s.
		In open black circles, the Expected Hierarchical Mutual Information (EHMI) averaged over the same pairs of partitions.
		In solid green squares, the ratio between the first and the second curves.
		In open blue squares, the EHMI for $\mathcal{T}=\mathcal{S}$ averaged over $\mathcal{T}$.
		Each point is averaged by sampling 1000 pairs of randomly generated hierarchical-partitions.
		The EHMI is computed by sampling permutations $\rho$ until the relative standard error of the mean falls below $0.01$.
	}
\end{figure}

\begin{figure}
	\includegraphics{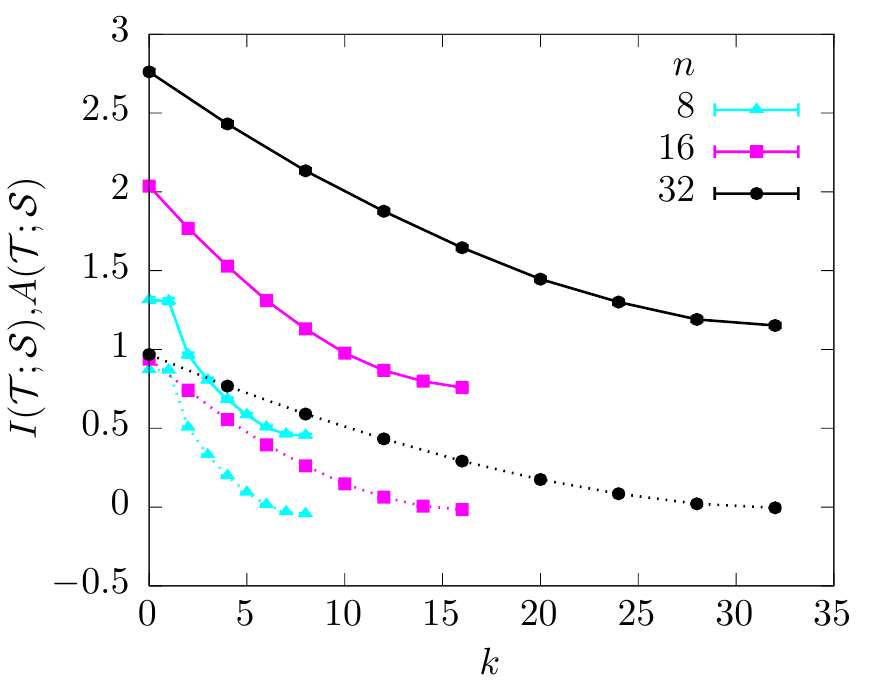}
	\caption{
		\label{fig:5}
		(Color online)
		Average Hierarchical Mutual Information $I$ (solid) and Adjusted Hierarchical Mutual Information $A$ (dotted) between randomly generated hierarchical-partitions $\mathcal{T}$ and corresponding hierarchical-partitions $\mathcal{S}$ obtained from $\mathcal{T}$ by randomly shuffling the identity of $k$ of the elements in $U$.
		Different symbols represent hierarchical-partitions of different sizes $n$.
		Each point is averaged over $10\,000$ samples of $\mathcal{T}$.
		The EHMI within the AHMI is computed as in Fig.~\ref{fig:4}.
	}
\end{figure}

In Fig.~\ref{fig:4} we show results concerning how similarities occurring by chance result in non-negligible values of the EHMI for randomly generated hierarchical-partitions.
The cyan curve of crosses depicts the average of the HMI between pairs of randomly generated hierarchical-partitions of $n$ elements.
In Appendix~\ref{app:E} we describe the algorithm we use to randomly sample hierarchical-partitions of $n$ elements.
The previous curve overlaps with the black one of open circles corresponding to the average of the EHMI between the same pairs of randomly generated hierarchical-partitions.
This result indicates that the permutation model is a good null model for the comparison of pairs of hierarchical-partitions without correlations.
Moreover, these curves exhibit significant positive values, indicating that the HMI detects similarities occurring just by chance between the randomly generated hierarchical-partitions. 
To determine how significant these values are, the curve of the magenta solid circles corresponds to the average of the hierarchical entropies of the generated hierarchical-partitions.
As can be seen, the averaged hierarchical entropy lies significantly above the curve of the EHMI.
On the other hand, their ratio, which is a quantity in $[0,1]$, is $\approx 0.3$ over the whole range of studied sizes, as indicated by the green curve of solid squares.
In other words, the similarities by chance affect non-negligibly the HMI.
The curve of open blue squares depicts the averaged EHMI but for $\mathcal{S}=\mathcal{T}$.
The curve lies above but follows closely that of the EHMI between different hierarchical-partitions.
This indicates that the effect of a randomized structure has a marginal impact besides that of the randomization of labels.

In Fig.~\ref{fig:5} we show how the HMI between two hierarchical-partitions $\mathcal{T}$ and $\mathcal{S}$ decays with $k$, when $\mathcal{S}$ is obtained from shuffling the identity of $k$ of the elements in $U$.
Here, the HMI is averaged by sampling randomly generated hierarchical-partitions $\mathcal{T}$ at each $n$ and $k$.
As expected, the average HMI decays as the imposed decorrelation increases.
In fact, for $k=n$ the obtained values match those of the EHMI (blue curve of open squares in Fig.~\ref{fig:4}).
In the figure, we also show the AHMI as a function of $k$ for the different $n$.
Notice how, at difference with the HMI, the AHMI goes from $A=1$ at $k=0$ to $A=0$ at $k=n$.

The previous results highlight the importance of the AHMI, in the sense that it conveys as a less biased measure of similarity as compared to the HMI.

\subsection{Example application\label{sec:III:C}}

Let us show a simple example application of the presented framework.
The small {\em animals} dataset~(see~\cite{kaufman1990finding}, Pag.~295) considers 6 boolean features for 20 rather arbitrarily selected animal species.
Within the 300 entries of the species-features boolean matrix, there are 5 missing or unspecified values.
In our example, we exploit the HMI and the HVI to infer the unspecified values and to analyze how the variation of these values affects the hierarchical classification of the species.

We generate $2^5=32$ variants of the species-features matrix by setting candidate values to the unspecified features.
From the matrices, we compute 32 corresponding hierarchical clusterings using the average-linkage clusterization algorithm equipped with the Manhattan distance~\cite{kaufman1990finding}.
Then, by removing the splitting distances, we convert the hierarchical clusterings into hierarchical partitions.
Here, non-binary partitions result from degenerate splitting distances.
The obtained ensemble of 32 hierarchical partitions embodies the uncertainty generated by the missing features.

The eccentricity of the $\alpha$-th hierarchical partition $\mathcal{T}_{\alpha}$ is defined by $C_{\alpha}:=(1/32)\sum_{\beta} V(\mathcal{T}_{\alpha};\mathcal{T}_{\beta})$, i.e. it is the average HVI between $\mathcal{T}_{\alpha}$ and the other hierarchical partitions in the ensemble.
The central hierarchical partition $\hat{\mathcal{T}}$ is the one minimizing the eccentricity, and it represents a parsimonious inference of the unspecified features.
The inference predicts that lobsters live in groups while frogs and salamanders do not, and that lions belong to an endangered species while spiders do not.
These are reasonable predictions.

\begin{figure}
    \hspace*{-1.0cm}
	\includegraphics[angle=0,scale=.5]{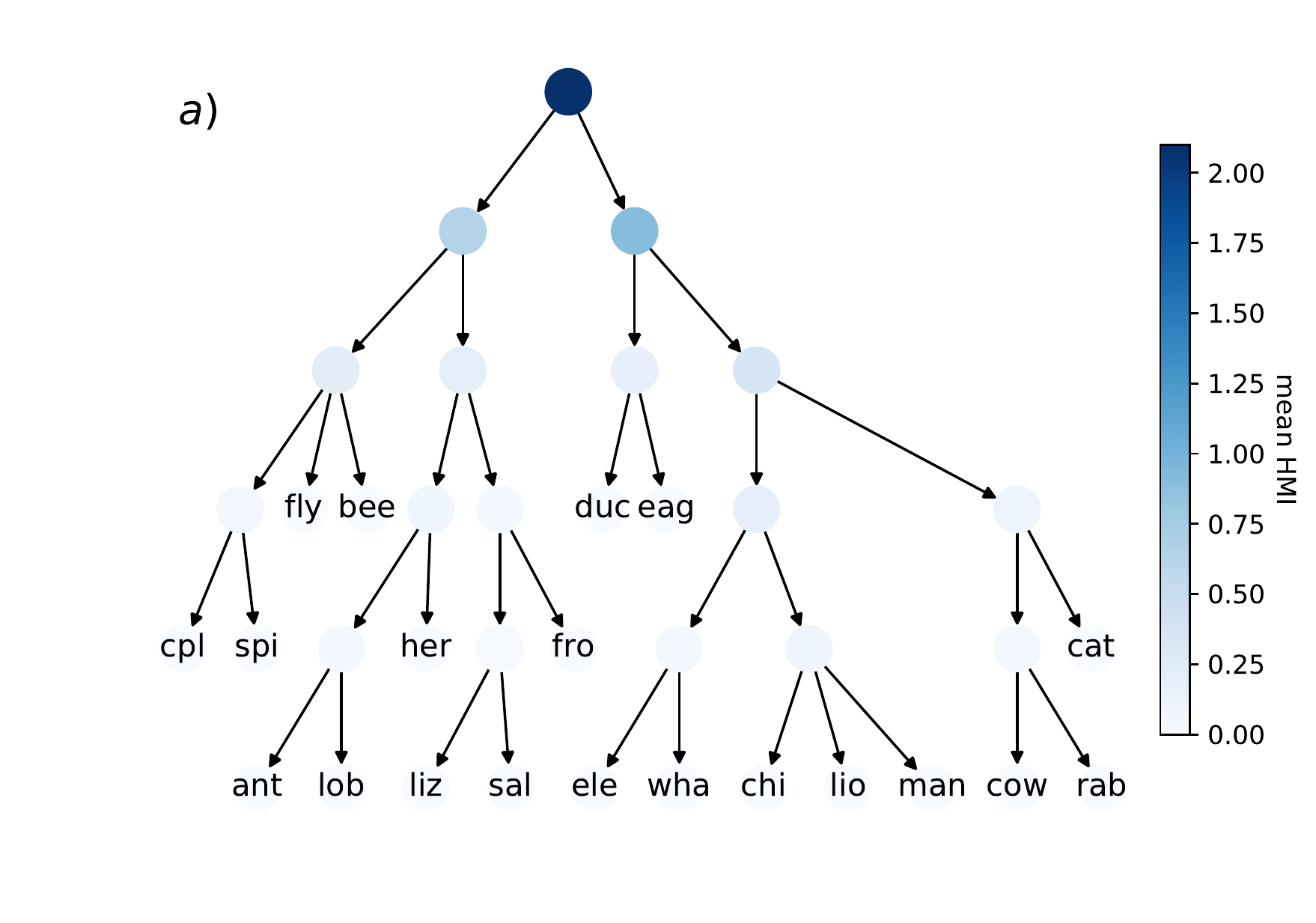}
    \vspace{-0.9cm}	
    \\
    \hspace*{-1.0cm}    
	\includegraphics[angle=0,scale=.5]{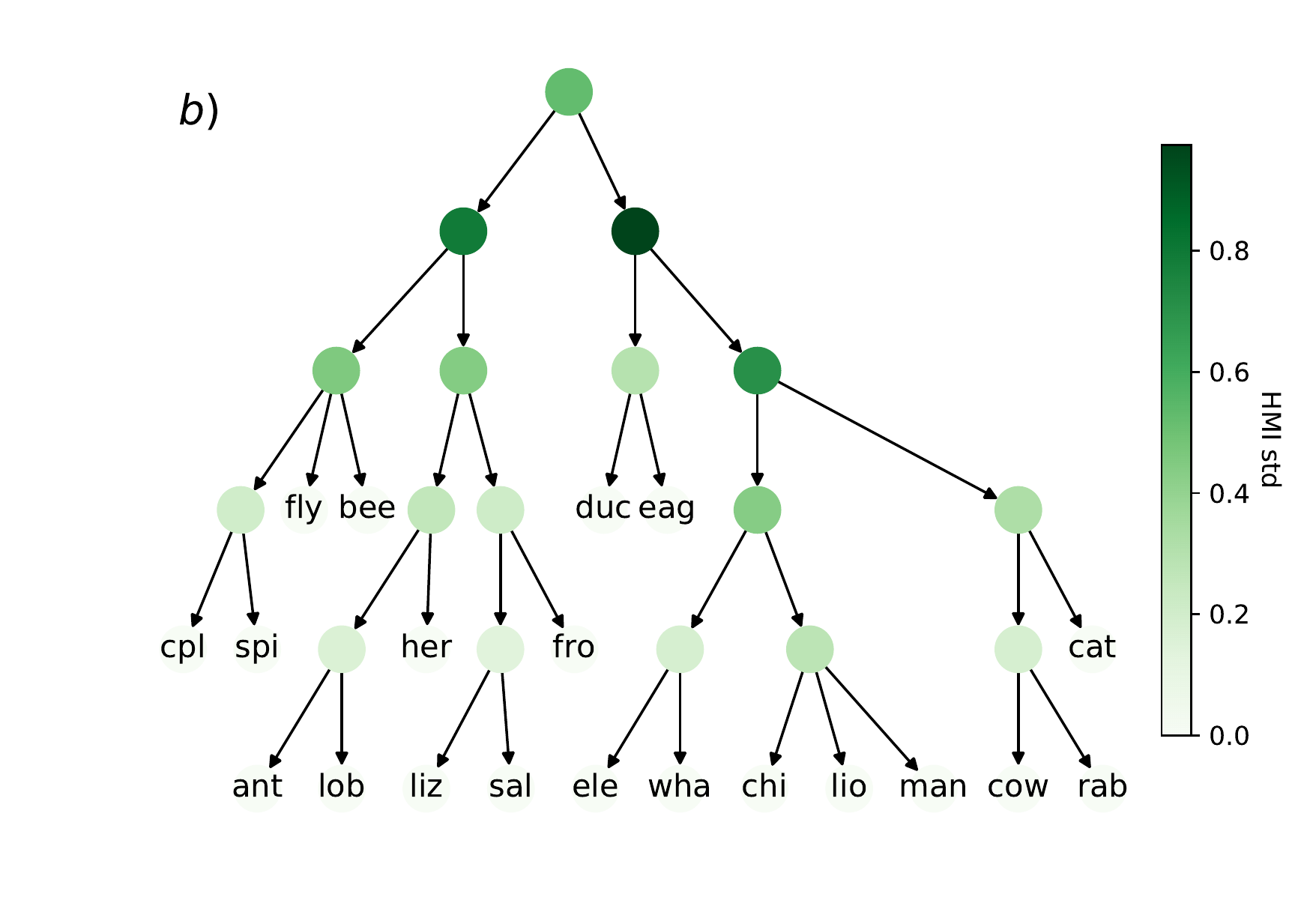}
    \vspace{-0.9cm}	
	\\
    \hspace*{-1.0cm}	
	\includegraphics[angle=0,scale=.5]{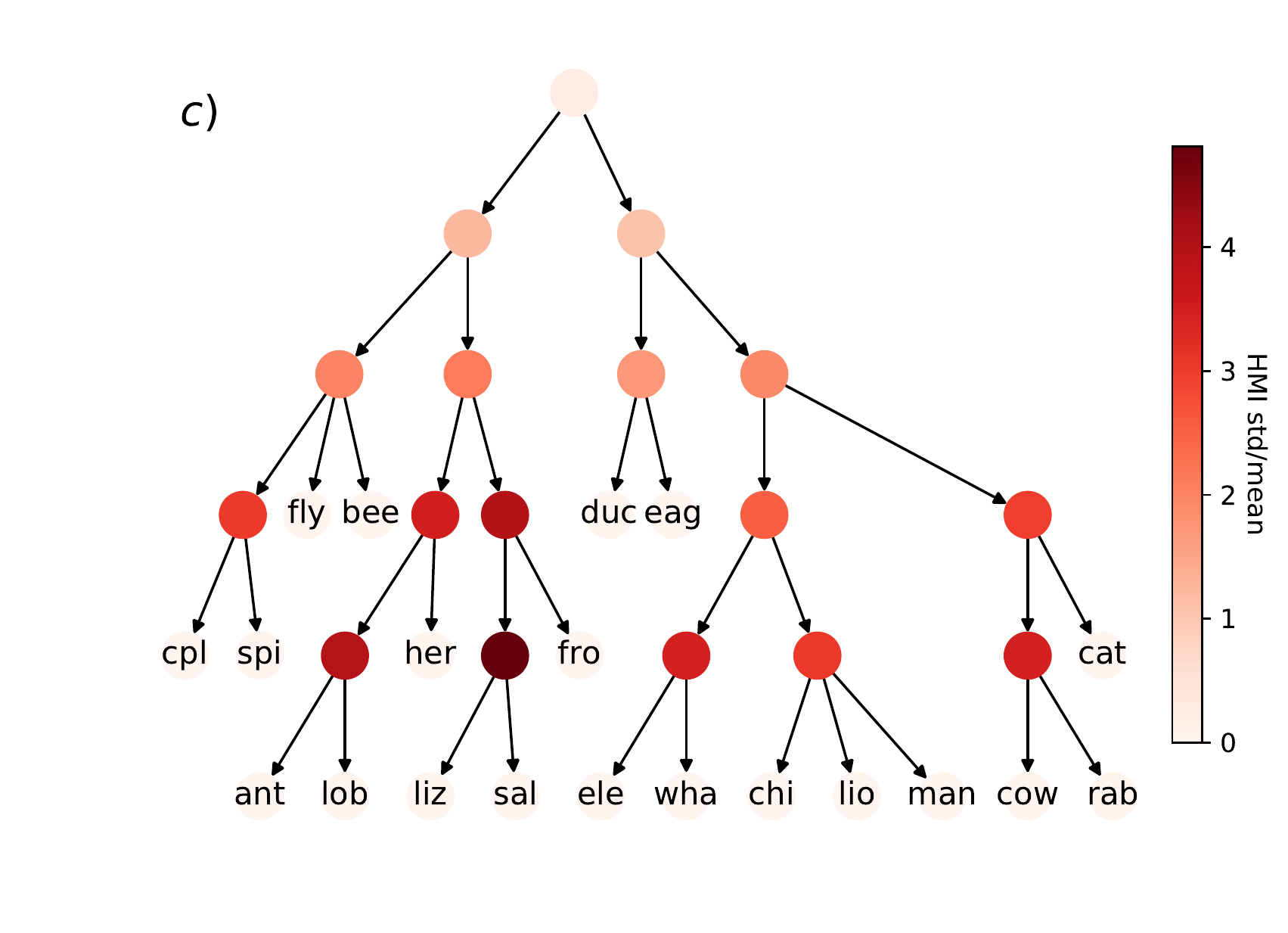}
    \vspace{-0.7cm}	
	\caption{
		\label{fig:6}
		(Color online)
        All panels show the most parsimonious hierarchical partition $\hat{\mathcal{T}}$ of 20 animal species classified by 6 boolean features.
        The most parsimonious hierarchical partition is selected among the $2^5=32$ candidates generated by varying the truth table of the 5 unspecified feature values found in the dataset. 
        All hierarchical partitions in the ensemble are obtained from corresponding hierarchical clusterings computed with the average-linkage algorithm with Manhattan distance.
        The selection minimizes the average HVI between the candidate hierarchical partition and the remaining ones.
        The color intensities indicate statistical values of the HMI between the vertices of $\hat{\mathcal{T}}$ and the vertices of all the hierarchical partitions in the ensemble.
        The mean in panel a), the standard deviation in b), and the standard deviation relative to the mean in c).
        As can be seen, uncertainty accumulates on top of the hierarchy since the vertices split large groups of species.
        Relative uncertainty accumulates on the bottom splits.    
	}
\end{figure}

To see how informative is each vertex of the most parsimonious hierarchical partition, we study the corresponding distribution of terms $I(\hat{\mathcal{T}}^t;\mathcal{T}_{\alpha}^s)$ (see Eq.~\ref{eq:2}) generated by the ensemble.
Here, we consider the different pairs of same level vertices $t$ and $s$ found in the hierarchical partitions $\hat{\mathcal{T}}$ and $\mathcal{T}_{\alpha}$, respectively, for the different $\alpha=1,...,32$.
From the distribution of values of $I(\hat{\mathcal{T}}^t;\mathcal{T}_{\alpha}^s)$, we compute three statistical estimators at each vertex $t$ of $\hat{\mathcal{T}}$.
The magnitudes of these values are depicted by the color intensities of Fig.~6. 
The mean is in Fig.~\ref{fig:6}a, the standard deviation in Fig.~\ref{fig:6}b, and the standard deviation relative to the mean in Fig.~\ref{fig:6}c.
The largest values of the mean and the standard deviation are found on the upper vertices since they correspond to the splitting of large groups of species, which produce large information gains.
On the other hand, the higher relative uncertainty is found in the vertices at the bottom (excepting leaves), since these vertices participate in the splitting of significantly different small groups of species as $\alpha$ varies.

\section{\label{sec:discussion} Discussion}

Our work shows that many similarities exist between classical information theory and the proposed generalization.
In this way, it significantly advances the generalization of information theory to hierarchical partitions.
We remark, however, that as with the quantum mechanical generalizations, significant differences also exist.
For instance, according to Eq.~\ref{eq:10}, multiple hierarchical partitions maximize the hierarchical entropy, as only the partition defined at the leaves contributes to the maximization, while the contribution of the internal levels produces no effect.
This result has relevant consequences.
For example, a straightforward generalization of the MaxEnt principle~\cite{cover2006elements} becomes ill-defined.
On the other hand, a slightly different reformulation of the principle solves the issue.
Namely, MaxEnt must be replaced by the maximization of the HMI with a reference hierarchical partition.
Since the classical MaxEnt is broadly applied in physics, our work can stimulate analogous contributions for the hierarchical case.
Another significant difference concerns the HVI.
Unlike its classical counterpart, we found that the HVI violates the triangular inequality.
On the other hand, we also found a transformation $d_n$ of the HVI satisfying the metric properties, consequently enabling the geometrization of the space of hierarchical partitions, although not in a universal way because the transformation is size-dependent.

Despite the significant contribution of our work, many important questions remain open for future investigation.
For instance, the cross-entropy plays an important role in the classical case.
It enables the definition of the information divergences, from where crucial results within classical information theory can be proven, such as the strong-additivity theorem~\cite{witten2020mini}.
Our work provides no hierarchical generalization of the cross-entropy, nor the divergences and the properties deriving from them.
It may be possible, however, a potentially equivalent proof of the monotonicity of the HMI.
Another open research question is the generalization of the HMI to the multivariate case.
Finally, how the generalization of information theory is related to encoding is also a topic for future research.
Progress on all these open issues must be achieved for the theory to mature.
\section{\label{sec:conclusions} Conclusions}

In several contexts of complex systems, information theory and statistical physics appear as an interwoven point of view, starting from the work of Jaynes~\cite{Jaynes1957}. 
Nevertheless, the study of an extension to hierarchical systems, while being crucial, for instance, for the comparison of various hierarchical structures, has been limited~\cite{perotti2015hierarchical}.
In this work, we proposed the generalization of information theory for hierarchical-partitions.
We analytically show that the Hierarchical Mutual Information (HMI) generalizes an important inequality of the classical non-hierarchical case.
We derive other information-theoretic quantities from the HMI: the Hierarchical Entropy, the Hierarchical Conditional Entropy, the Hierarchical Variation of Information (HVI), and the Adjusted Hierarchical Mutual Information (AHMI).
We studied the metric properties of the HVI, finding counter-examples violating the triangular inequality, and thus showing that the HVI fails to have the metric property of its non-hierarchical analogous.
On the other hand, we found a transformation $d_n$ of the HVI satisfying the metric properties, and thus enabling a geometrization of the space of hierarchical partitions.
Additionally, we supported the analytical findings with corresponding numerical experiments and an illustrative application with the hierarchical clustering of animal species.
We offer open-source access to our code~\cite{perotti2019hit_gitcode}, including the code for the generation of hierarchical-partitions.

Our work opens new possibilities in the study of hierarchically organized physical systems, from the information-theoretic side, the statistical side, as well as from the applications point of view.
From the theoretical point of view, we outlined several topics for future research that could further contribute to the development of the generalization of information theory for hierarchical partitions.
For instance, future studies may consider to incorporate a multivariate extension, the hierarchical cross-entropy, and the generalization of related divergences.
From the statistical point of view, future research may consider the generalization of the MaxEnt principle to the hierarchical case.
Finally, from the application point of view, it would be interesting to perform a comparative analysis including the information-theoretic metrics or, among similar possibilities, to use them to compute consensus taxonomic and phylogenetic trees~\cite{miralles2013metrics,salichos2014novel}.

\section{Acknowledgments}
\label{sec:acknowledgments}

JIP and NA acknowledge financial support from grants CONICET (PIP 112 20150 10028), FonCyT (PICT-2017-0973), SeCyT–UNC (Argentina) and MinCyT C\'ordoba (PID PGC 2018).
FS acknowledges support from the European Project SoBigData++ GA. 871042 and the PAI (Progetto di Attivit\`a Integrata) project funded by the IMT School Of Advanced Studies Lucca.
The authors thank CCAD -- Universidad Nacional de C\'ordoba, \url{http://ccad.unc.edu.ar/}, which is part of SNCAD -- MinCyT, Argentina, for the provided computational resources.


%

\onecolumngrid

\appendix

\section{Rewriting the HMI}
\label{app:A}

It is convenient to begin rewriting the hierarchical mutual information in the following alternative form, which is more convenient for our purposes
\begin{eqnarray}
\label{eq:A1}
I(\mathcal{T};\mathcal{S})
&=&
I(\mathcal{T}^{t_0};\mathcal{S}^{s_0})
\\
&=&
I(T_{t_0};S_{s_0}|t_0 s_0)
\nn
&&
+
\sum_{t_1\in T_{t_0},s_1\in S_{s_0}}
P(t_1 s_1|t_0 s_0)
\bigg\{
I(T_{t_1};S_{s_1}|t_1 s_1)
+
\sum_{t_2\in T_{t_1},s_2\in S_{s_1}}
P(t_2 s_2|t_1 s_1)
I(\mathcal{T}_{t_2};\mathcal{S}_{s_2})
\bigg\}
\nn
&=&
I(T_{t_0};S_{s_0}|t_0 s_0)
\nn
&&
+
\sum_{t_1\in T_1,s_1\in S_1}
P(t_1 s_1|t_0 s_0)
I(T_{t_1};S_{s_1}|t_1 s_1)
+
\sum_{t_2\in T_2,s_2\in S_2}
P(t_2 s_2|t_0 s_0)
I(\mathcal{T}_{t_2};\mathcal{S}_{s_2})
\nn
&\vdots&
\nn
&=&
\sum_{\ell=0}^{L-1}
\sum_{t\in T_{\ell},s\in S_{\ell}}
P(t s)
I(T_t;S_s|t s)
.
\nnn
\end{eqnarray}
Here, we used the definition $P(t s):=P(t s|t_0 s_0)=|U_t\cap U_s|/|U|=n_{ts}/n$.
Similarly
\begin{eqnarray}
\label{eq:A2}
\sum_{t\in T_{\ell},s\in S_{\ell}}
P(t s)
H(T_t,S_s|t s)
&=&
\sum_{t\in T_{\ell},s\in S_{\ell}}
P(t s)
\sum_{t'\in T_t,s'\in S_s}
-
P(t' s'|t s)
\ln
P(t' s'|t s)
\\
&=&
\sum_{t\in T_{\ell},s\in S_{\ell}}
\sum_{t'\in T_t,s'\in S_s}
-
P(t' s' t s)
\bigg[
\ln
P(t' s' t s)
-
\ln
P(t s)
\bigg]
\nn
&=&
\sum_{t\in T_{\ell},s\in S_{\ell}}
\sum_{t'\in T_{\ell+1},s'\in S_{\ell+1}}
-
P(t' s' t s)
\ln
P(t' s' t s)
\nn
&&
-
\sum_{t\in T_{\ell},s\in S_{\ell}}
\sum_{t'\in T_{\ell+1},s'\in S_{\ell+1}}
-
P(t' s' t s)
\ln
P(t s)
\nn
&=&
\sum_{t\in T_{\ell},s\in S_{\ell}}
\sum_{t'\in T_{\ell+1},s'\in S_{\ell+1}}
-
P(t' s' t s)
\ln
P(t' s' t s)
\nn
&&
-
\sum_{t\in T_{\ell},s\in S_{\ell}}
-
P(t s)
\ln
P(t s)
\nn
&=&
H(T_{\ell+1},S_{\ell+1},T_{\ell},S_{\ell})
-
H(T_{\ell},S_{\ell})
\nn
&=&
H(T_{\ell+1},S_{\ell+1}|T_{\ell},S_{\ell}).
\nnn
\end{eqnarray}
where we used that
$\sum_{t\in T_{\ell}}\sum_{t'\in T_{\ell+1}} 
\equiv 
\sum_{t\in T_{\ell}}(\sum_{t'\in T_t}+\sum_{t'\in T_{\ell+1}/T_t}) 
\equiv 
\sum_{t\in T_{\ell}}(\sum_{t'\in T_t}+0)$
because $P(t's'ts)=0$ whenever $t'$ is not a child of $t$.
The entropies in the last two lines are written in terms of the standard non-hierarchical or classical definition, for which
\begin{eqnarray}
\label{eq:A3}
H(X',Y'|X,Y)
=
\sum_{x\in X,y\in Y}
P(xy)
\sum_{x'\in X',y'\in Y'}
-
P(x'y'|xy)
\ln
P(x'y'|xy)
\end{eqnarray}
Finally, combining Eqs.~\ref{eq:A1}~and~\ref{eq:A2} we arrive at
\begin{eqnarray}
\label{eq:A4}
I(\mathcal{T};\mathcal{S})
&=&
\sum_{\ell=0}^{L-1}
\sum_{t\in T_{\ell},s\in S_{\ell}}
P(t s)
I(T_t;S_s|t s)
\\
&=&
\sum_{\ell=0}^{L-1}
\sum_{t\in T_{\ell},s\in S_{\ell}}
P(t s)
\bigg[
H(T_t|t s )
+
H(S_s|t s)
-
H(T_t,S_s|t s)
\bigg]
\nn
&=&
\sum_{\ell=0}^{L-1}
\bigg[
H(T_{{\ell}+1}|T_{\ell}, S_{\ell})
+
H(S_{{\ell}+1}|T_{\ell}, S_{\ell})
-
H(T_{{\ell}+1},S_{{\ell}+1}|T_{\ell},S_{\ell})
\bigg]
\nn
&=&
\sum_{\ell=0}^{L-1}
I(T_{{\ell}+1};S_{{\ell}+1}|T_{\ell},S_{\ell})
.
\nnn
\end{eqnarray}

\section{HMI inequality}
\label{app:B}

The first inequality in Eq.~\ref{eq:8} follows because
$I(T_{{\ell}+1};S_{{\ell}+1}|T_{\ell},S_{\ell}) \geq 0$ for any $\ell$.
For the second inequality, we start from Eq.~\ref{eq:A1}.
We can write
\begin{eqnarray}
\label{eq:B1}
I(\mathcal{T};\mathcal{S})
&=&
\sum_{\ell=0}^{L-1}
\sum_{t\in T_{\ell},s\in S_{\ell}}
P(ts)
I(T_t;S_s|ts)
\\
&\leq &
\sum_{\ell=0}^{L-1}
\sum_{t\in T_{\ell},s\in S_{\ell}}
P(ts)
H(T_t|ts)
\nn
&=&
\sum_{\ell=0}^{L-1}
\sum_{t\in T_{\ell},s\in S_{\ell}}
P(ts)
\sum_{t'\in T_t}
-
P(t'|ts)
\ln
P(t'|ts)
\nn
&=&
\sum_{\ell=0}^{L-1}
\sum_{s\in S_{\ell}}
\sum_{t\in T_{\ell}}
\sum_{t'\in T_t}
-
P(t's)
\ln
P(t'|ts)
\nn
&\leq &
\sum_{\ell=0}^{L-1}
\sum_{t\in T_{\ell}}
\sum_{t'\in T_t}
-
\left(
\sum_{s\in S_{\ell}}
P(t's)
\right)
\ln
\left(
\frac{
	\sum_{s\in S_{\ell}}
	P(t's)
}{
	\sum_{s\in S_{\ell}}
	P(ts)
}
\right)
\nn
&= &
\sum_{\ell=0}^{L-1}
\sum_{t\in T_{\ell}}
P(t)
\sum_{t'\in T_t}
-
P(t'|t)
\ln
P(t'|t)
\nn
&= &
\sum_{\ell=0}^{L-1}
\sum_{t\in T_{\ell}}
P(t)
H(T_t|t)
\nn
&= &
\sum_{\ell=0}^{L-1}
H(T_{\ell+1}|T_{\ell}) 
\nn
&= &
\sum_{\ell=0}^{L-1}
\bigg[
H(T_{\ell+1}|T_{\ell},T_{\ell})
+
H(T_{\ell+1}|T_{\ell},T_{\ell})
-
H(T_{\ell+1},T_{\ell+1}|T_{\ell},T_{\ell})
\bigg]
\nn
&= &
\sum_{\ell=0}^{L-1}
I(T_{\ell+1};T_{\ell+1}|T_{\ell},T_{\ell})
\nn
&= &
I(\mathcal{T};\mathcal{T})
.
\nnn
\end{eqnarray}
Here, in the first inequality, we used a well-known property of the entropy, while in the second inequality, we used the log-sum inequality~\cite{cover2006elements}.

\section{Generating hierarchical-partitions}
\label{app:C}

Before showing how to generate all hierarchical-partitions of a set, let us first review a way to generate all standard partitions (see Section 7.2.1.7 of~\cite{knuth2011art}).
Consider we have a way to generate all partitions of the set $U_n:=\{1,2,...,n\}$.
Then, we can easily generate all the partitions of the set
$U_{n+1}=\{1,2,...,n,n+1\}$ as follows.
For each partition of the set $U_n$, generate all the partitions that can be obtained by adding the element $n+1$ to each part $\mathcal{P}$ together with extending the partition with the part $\{n+1\}$.
For example, given the partition
$\{\{1,2\},\{3\}\}$ of $\{1,2,3\}$, then we generate the partitions
$\{\{1,2,4\},\{3\}\}$,
$\{\{1,2\},\{3,4\}\}$
and
$\{\{1,2\},\{3\},\{4\}\}$
of
$\{1,2,3,4\}$.
In other words, this algorithm recursively implements induction.

To generate hierarchical-partitions, we follow a similar procedure to the one discussed for standard partitions.
Consider we have an algorithm to generate all hierarchical-partitions of $U_n$.
Then, for each hierarchical-partition $\mathcal{T}$ of $U_n$, we generate the hierarchical-partitions $\mathcal{T}'$ of $U_{n+1}$ that can be obtained by applying the following operations to each of the nodes $t\in \mathcal{T}$:
\begin{enumerate}
	\item If $t$ is a leaf, add $n+1$ to $U_t$.
	\item If $t$ is not a leaf, add the child $t'$ to $t$ with $U_{t'}=\{n+1\}$.
	\item Replace $t$ by a new node $t''$ with $t$ and $t'$ as children.
\end{enumerate}
For example, the hierarchical-partitions of $U_2=\{1,2\}$ are
$\{1,2\}$ and $\{\{1\},\{2\}\}$. 
Then, the following applies.

Operation 1 applied to the first hierarchical-partition results in $\{1,2,3\}$.
Operation 1 applied to the second results in 
$\{\{1,3\},\{2\}\}$ and $\{\{1\},\{2,3\}\}$.
Operation 2 on the second, results in the hierarchical-partitions $\{\{1\},\{2\},\{3\}\}$.
Operation 3 on the first, results in 
$\{\{1,2\},\{3\}\}$.
Operation 3 on the second, results in 
$\{\{\{1\},\{2\}\},\{3\}\}$,
$\{\{\{1\},\{3\}\},\{2\}\}$ and
$\{\{1\},\{\{2\},\{3\}\}\}$.
For more details, please check our code for an implementation of the algorithm~\cite{perotti2019hit_gitcode}.

\section{Forcing triangular inequality for the Hierarchical Variation of Information}
\label{app:D}

Let 
\begin{eqnarray}
\label{eq:D1}
d_{V_0}(\mathcal{T};\mathcal{S})
&:=&
1-e^{-V(\mathcal{T};\mathcal{S})/V_0}
\end{eqnarray}
be defined for some arbitrary $V_0>0$.
Then, for an appropriate choice of $V_0$, $d_{V_0}$ becomes a distance metric satisfying the triangular inequality.
The proof is as follows.
First, $d_{V_0}$ is clearly a distance since: {\em i)} $d_{V_0}$ is a growing function of $V$, {\em ii)} $d_{V_0}(\mathcal{T},\mathcal{S})=0 \Leftrightarrow \mathcal{T} = \mathcal{S}$ when $V_0>0$ and {\em iii)} $d_{V_0}$ is symmetric in its arguments.
It remains to be shown that $d_{V_0}$ satisfies the triangular inequality for an appropriate choice of $V_0$.
The triangular inequality for $d_{V_0}$ reads
\begin{eqnarray}
\label{eq:D2}
\Delta d_{V_0}(\mathcal{T};\mathcal{S};\mathcal{R})
&:=&
d_{V_0}(\mathcal{T};\mathcal{S})
+
d_{V_0}(\mathcal{S};\mathcal{R})
-
d_{V_0}(\mathcal{T};\mathcal{R})
\\
&\geq&
1
-e^{-V(\mathcal{T};\mathcal{S})/V_0}
-e^{-V(\mathcal{S};\mathcal{R})/V_0}
\nn
&\geq&
1
-
2e^{-\min\{V(\mathcal{T};\mathcal{S}),V(\mathcal{S};\mathcal{R})\}/V_0}
.
\nnn
\end{eqnarray}
We can show that, for an appropriate choice of $V_0$, last line is always non-negative, given that non-zero values of $V$ cannot be arbitrarily small.
Thus, let us find a lower bound for the non-zero values of the Variation of Information between hierarchical-partitions.
To do so, first, we notice that the Variation of Information between hierarchical-partitions can be decomposed into a summation of non-negative quantities over the different levels.
Namely, following Eqs.~\ref{eq:7},~\ref{eq:10}~and~\ref{eq:13}, we can write
\begin{eqnarray}
\label{eq:D3}
V(\mathcal{T};\mathcal{S})
&=&
\sum_{\ell=0}^{L-1}
\bigg[
H(T_{\ell+1}|T_{\ell})
+
H(S_{\ell+1}|S_{\ell})
-
2I(T_{\ell+1};S_{\ell+1}|T_{\ell},S_{\ell})
\bigg]
\\
&=:&
\sum_{\ell=0}^{L-1} 
V(T_{\ell+1};S_{\ell+1}|T_{\ell},S_{\ell})
\nnn
\end{eqnarray}
with $V(T_{\ell+1};S_{\ell+1}|T_{\ell},S_{\ell}) \geq 0$ for every $\ell$ due to Eq.~\ref{eq:8}.
Now, if the hierarchical-partitions 
$\mathcal{T}$ and $\mathcal{S}$
are equal up to level $\ell'$ included (i.e., as stochastic variables, $T_{\ell}=S_{\ell}$ for all $\ell\leq \ell'$) then
\begin{eqnarray}
\label{eq:D4}
I(\mathcal{T};\mathcal{S})
&=&
\sum_{\ell=0}^{\ell'}
I(T_{\ell+1};S_{\ell+1}|T_{\ell},S_{\ell})
+
\sum_{\ell=\ell'+1}^{L-1}
I(T_{\ell+1};S_{\ell+1}|T_{\ell},S_{\ell})
\\
&=&
I(T_{\ell'+1};S_{\ell'+1}|T_0,S_0)
+
\sum_{\ell=\ell'+1}^{L-1}
I(T_{\ell+1};S_{\ell+1}|T_{\ell},S_{\ell})
\nnn
\end{eqnarray}
because
\begin{eqnarray}
\label{eq:D5}
\sum_{\ell=0}^{\ell'}
I(T_{\ell+1};S_{\ell+1}|T_{\ell},S_{\ell})
&=&
I(T_{\ell'+1};S_{\ell'+1}|T_{\ell'},S_{\ell'})
+
\sum_{\ell=0}^{\ell'-1}
I(T_{\ell+1};S_{\ell+1}|T_{\ell},S_{\ell})
\\
&=&
I(T_{\ell'+1};S_{\ell'+1}|T_{\ell'},S_{\ell'})
+
\sum_{\ell=0}^{\ell'-1}
H(T_{\ell+1}|T_{\ell})
\nn
&=&
H(T_{\ell'+1}|T_{\ell'})
+
H(S_{\ell'+1}|S_{\ell'})
-
H(T_{\ell'+1},S_{\ell'+1}|S_{\ell'})
+
\sum_{\ell=0}^{\ell'-1}
\bigg[
H(T_{\ell+1},T_{\ell})-H(T_{\ell})
\bigg]
\nn
&=&
H(T_{\ell'+1},T_{\ell'})
-
H(T_{\ell'})
+
H(S_{\ell'+1},S_{\ell'})
-
H(S_{\ell'})
-
H(T_{\ell'+1},S_{\ell'+1},T_{\ell'},S_{\ell'})
+
H(T_{\ell'},S_{\ell'})
\nn
&&
+
\sum_{\ell=0}^{\ell'-1}
\bigg[
H(T_{\ell+1},T_{\ell})-H(T_{\ell})
\bigg]
\nn
&=&
H(T_{\ell'+1})
-
H(T_{\ell'})
+
H(S_{\ell'+1})
-
H(S_{\ell'})
-
H(T_{\ell'+1},S_{\ell'+1})
+
H(S_{\ell'})
+
H(T_{\ell'})-H(T_{0})
\nn
&=&
H(T_{\ell'+1})
+
H(S_{\ell'+1})
-
H(T_{\ell'+1},S_{\ell'+1})
-
0
\nn
&=&
I(T_{\ell'+1};S_{\ell'+1}|T_{0},S_{0}).
\nnn
\end{eqnarray}
Here, we used identities such as 
\begin{eqnarray}
\label{eq:D6}
H(T_{\ell+1},T_{\ell})
&=&
\sum_{t\in T_{\ell}}
\sum_{t'\in T_{\ell+1}}
-
P(t't) \ln P(t't)
\\
&=&
-
\sum_{t\in T_{\ell}}
\bigg(
\sum_{t'\in T_{t}}
P(t't) \ln P(t't)
+
\sum_{t'\in T_{\ell+1}/T_{t}}
P(t't) \ln P(t't)
\bigg)
\nn
&=&
-
\sum_{t\in T_{\ell}}
\bigg(
\sum_{t'\in T_{t}}
P(t't) \ln P(t't)
+
0
\bigg)
,
\;\;\;\;
\mbox{because if $t'\notin T_t$ then $P(t't)=0$,}
\nn
&=&
-
\sum_{t\in T_{\ell}}
\sum_{t'\in T_{t}}
P(t't) \ln P(t't)
\nn
&=&
-
\sum_{t\in T_{\ell}}
\sum_{t'\in T_{t}}
P(t't) \ln P(t')
,
\;\;\;\;
\mbox{because $P(t't)=P(t')$ since $U_{t'}\subseteq U_t$ whenever $t'\in T_t$,}
\nn
&=&
-
\sum_{t\in T_{\ell}}
\sum_{t'\in T_{\ell+1}}
P(t't) \ln P(t')
,
\;\;\;\;
\mbox{because of the same trick as in the second and third lines,}
\nn
&=&
-
\sum_{t'\in T_{\ell+1}}
\ln P(t')
\sum_{t\in T_{\ell}}
P(t't)
\nn
&=&
-
\sum_{t'\in T_{\ell+1}}
P(t')
\ln P(t')
\nn
&=&
H(T_{\ell+1})
\nnn
\end{eqnarray}
and
$H(T_{\ell'},S_{\ell'})=H(S_{\ell'})$.
Combining Eqs.~\ref{eq:D3}~and~\ref{eq:D5} we can write
\begin{eqnarray}
\label{eq:D7}
V(\mathcal{T};\mathcal{S})
&=&
V(T_{\ell'+1};S_{\ell'+1}|T_0,S_0)
+
\sum_{\ell=\ell'+1}^{L-1}
V(T_{\ell+1};S_{\ell+1}|T_{\ell},S_{\ell})
.
\end{eqnarray}
Now, as shown in Ref.~\cite{meila2007comparing}, the Variation of Information between two different classical partitions cannot be smaller than $2/n$ when the size of the universe is $n=|U|$.
In consequence, since $T_0=S_0=U$, then 
$V(\mathcal{T};\mathcal{S}) \geq V(T_{\ell'+1};S_{\ell'+1}|T_0,S_0) \geq 2/n$.
Finally, from this lower bound and Eq.~\ref{eq:D2} we have
$\Delta d_{V_0}(\mathcal{T};\mathcal{S};\mathcal{R})
\geq 
1-2e^{-2/(nV_0)}$
from where, by setting the right-hand side (r.h.s.) to zero, we obtain $V_0=2/(n\ln 2)$.
In other words, we showed that
\begin{eqnarray}
\label{eq:D8}
d_n(\mathcal{T};\mathcal{S})
:=
1-e^{-n\tfrac{\ln 2}{2}V(\mathcal{T};\mathcal{S})}
\end{eqnarray}
satisfies the triangular inequality and thus is a distance metric with image in $[0,d_{\max}]$ with $d_{\max}\lesssim 1$.

\section{Generating random hierarchical-partitions}
\label{app:E}

To generate or sample random hierarchical-partitions in a non-necessarily uniform manner we propose a recursive application of an algorithm to generate random partitions from a set of elements $U$.

To generate random partitions of a set $U$ of $n$ elements, we first draw a number $z$ of ``splitters''  uniformly at random from the set $\{0,1,2,...,n\}$.
Then, we generate a sequence concatenating the $z$ splitters $|$ with the $n$ elements of $U$.
Then, we randomly shuffle the sequence.
Then, we split the sequence by removing the splitters and use the obtained non-empty parts to construct a partition.
For example, if $U=\{1,2,3,4,5\}$ and $z=3$, then we generate the sequence $|||12345$ which after shuffling may  result in $12|3||45$ from where the partition $\{\{1,2\},\{3\},\{4,5\}\}$ is obtained.

To generate random hierarchical-partitions, we recursively apply the previous algorithm, first to $U$, then to the obtained parts of $U$, then to the parts of the parts, and so on until non-divisible sets are obtained.
For details please check our code~\cite{perotti2019hit_gitcode}.

\end{document}